# Evaluation of E-Learners Behaviour using Different Fuzzy Clustering Models: A Comparative Study

Mofreh A. Hogo*
Dept. of Electrical Engineering Technology, Higher Institution of Technology Benha, Benha University, Egypt.

*Abstract*— **This paper introduces an evaluation methodologies for the e-learners' behaviour that will be a feedback to the decision makers in e-learning system. Learner's profile plays a crucial role in the evaluation process to improve the e-learning process performance. The work focuses on the clustering of the e-learners based on their behaviour into specific categories that represent the learner's profiles. The learners' classes named as regular, workers, casual, bad, and absent. The work may answer the question of how to return bad students to be regular ones. The work presented the use of different fuzzy clustering techniques as fuzzy c-means and kernelized fuzzy c-means to find the learners' categories and predict their profiles. The paper presents the main phases as data description, preparation, features selection, and the experiments design using different fuzzy clustering models. Analysis of the obtained results and comparison with the real world behavior of those learners proved that there is a match with percentage of 78%. Fuzzy clustering reflects the learners' behavior more than crisp clustering. Comparison between FCM and KFCM proved that the KFCM is much better than FCM in predicting the learners' behaviour.**

## I. INTRODUCTION

The development of web-based education systems have grown exponentially in the last years [1]. These systems accumulate a great deal of information; which is very valuable in analyzing students' behavior and assisting teachers in the detection of possible errors, shortcomings and improvements. However, due to the vast quantities of data these systems can generate daily, it is very difficult to manage manually, and authors demand tools which assist them in this task, preferably on a continuous basis. The use of data mining is a promising area in the achievement of this objective [2]. In the knowledge discovery in databases (KDD) process, the data mining step consists of the automatic extraction of implicit and interesting patterns from large data collections. A list of data mining techniques or tasks includes statistics, clustering, classification, outlier detection, association rule mining, sequential pattern mining, text mining, or subgroup discovery, among others [3]. In recent years, researchers have begun to investigate various data mining methods in order to help teachers improve e-learning systems. A review can be seen in [2]; these methods allow the discovery of new knowledge based on students' usage data. Subgroup discovery is a specific method for discovering descriptive rules [4,5].

## II. SURVEY ON E-LEARNING

### A. Clustering

The first application of clustering methods in e-learning [6], a network-based testing and diagnostic system was implemented. It entails a multiple-criteria test-sheet-generating problem and a dynamic programming approach to generate test sheets. The proposed approach employs fuzzy logic theory to determine the difficulty levels of test items according to the learning status and personal features of each student, and then applies an Artificial Neural Network model: Fuzzy Adaptive Resonance Theory (Fuzzy ART) [7] to cluster the test items into groups, as well as dynamic programming [8] for test sheet construction. In [9], an in-depth study describing the usability of Artificial Neural Networks and, more specifically, of Kohonen's Self-Organizing Maps (SOM) [10] for the evaluation of students in a tutorial supervisor (TS) system, as well as the ability of a fuzzy TS to adapt question difficulty in the evaluation process, was carried out. An investigation on how Data Mining techniques could be successfully incorporated to e-learning environments, and how this could improve the learning processes was presented in [11]. Here, data clustering is suggested as a means to promote group-based collaborative learning and to provide incremental student diagnosis. In [12], user actions associated to students' Web usage were gathered and preprocessed as part of a Data Mining process. The Expectation Maximization (EM) algorithm was then used to group the users into clusters according to their behaviors. These results could be used by teachers to provide specialized advice to students belonging to each cluster. The simplifying assumption that students belonging to each cluster should share Web usage behavior makes personalization strategies more scalable. The system administrators could also benefit from this acquired knowledge by adjusting the e-learning environment they manage according to it. The EM algorithm was also the method of choice in [13], where clustering was used to discover user behavior patterns in collaborative activities in e-

(IJCSIS) International Journal of Computer Science and Information Security,
Vol. 7, No. 2, 2010

learning applications. Some researchers [14-16], propose the use of clustering techniques to group similar course materials: An ontology-based tool, within a Web Semantics framework, was implemented in [16] with the goal of helping e-learning users to find and organize distributed courseware resources. An element of this tool was the implementation of the Bisection K-Means algorithm, used for the grouping of similar learning materials. Kohonen's well-known SOM algorithm was used in [14] to devise an intelligent searching tool to cluster similar learning material into classes, based on its semantic similarities. Clustering was proposed in [15] to group similar learning documents based on their topics and similarities. A Document Index Graph (DIG) for document representation was introduced, and some classical clustering algorithms (Hierarchical Agglomerative Clustering, Single Pass Clustering and k-NN) were implemented. Different variants of the Generative Topographic Mapping (GTM) model, a probabilistic alternative to SOM, were used in [17-19] for the clustering and visualization of multivariate data concerning the behavior of the students of a virtual course. More specifically, in [17, 18] a variant of GTM known to behave robustly in the presence of atypical data or outliers was used to successfully identify clusters of students with atypical learning behaviors. A different variant of GTM for feature relevance determination was used in [19] to rank the available data features according to their relevance for the definition of student clusters.

### B. Prediction Techniques

The forecasting of students' behavior and performance when using e-learning systems bears the potential of facilitating the improvement of virtual courses as well as e-learning environments in general. A methodology to improve the performance of developed courses through adaptation was presented in [20,21]. Course log-files stored in databases could be mined by teachers using evolutionary algorithms to discover important relationships and patterns, with the target of discovering relationships between students' knowledge levels, e-learning system usage times and students' scores. A system for the automatic analysis of user actions in Web-based learning environments, which could be used to make predictions on future uses of the learning environment, was presented in [22]. It applies a C4.5 DT model for the analysis of the data; (Note that this reference could also have been included in the section reviewing classification methods). Some studies apply regression methods for prediction [23-25]. In [24], a study that aimed to find the sources of error in the prediction of students' knowledge behavior was carried out. Stepwise regression was applied to assess what metrics help to explain poor prediction of state exam scores. Linear regression was applied in [25] to predict whether the student's next response would be correct, and how long he or she would take to generate that response. In [25], a set of experiments was conducted in order to predict the students' performance in e-learning courses, as well as to assess the relevance of the attributes involved. In this approach, several Data Mining methods were applied, including: Naïve Bayes, KNN, MLP Neural Network, C4.5, Logistic Regression, and Support Vector Machines. Rule extraction was also used in [20,21] with the emphasis on the discovery of interesting prediction rules in student usage information, in order to use them to improve adaptive Web courses. Graphical models and Bayesian methods have also been used in this context. Some models for the detection of atypical student behavior were also referenced in the section reviewing clustering applications [17,19].

### C. Fuzzy Logic-Based Methods

These methods have only recently taken their first steps in the e-learning field [26-28]. For example in, [28] a Neurofuzzy model for the evaluation of students in an intelligent tutoring system (ITS) was presented. Fuzzy theory was used to measure and transform the interaction between the student and the ITS into linguistic terms. Then, Artificial Neural Networks were trained to realize fuzzy relations operated with the max–min composition. These fuzzy relations represent the estimation made by human tutors of the degree of association between an observed response and a student characteristic. A fuzzy group-decision approach to assist users and domain experts in the evaluation of educational Web sites was realized in the EWSE system, presented in [27]. In further work by Hwang and colleagues [26,27], a fuzzy rules-based method for eliciting and integrating system management knowledge was proposed and served as the basis for the design of an intelligent management system for monitoring educational Web servers. This system is capable of predicting and handling possible failures of educational Web servers, improving their stability and reliability. It assists students' self-assessment and provides them with suggestions based on fuzzy reasoning techniques. A two-phase fuzzy mining and learning algorithm was described in [27]. It integrates an association rule mining algorithm, called Apriori, with fuzzy set theory to find embedded information that could be fed back to teachers for refining or reorganizing the teaching materials and tests. In a second phase, it uses an inductive learning algorithm of the AQ family: AQR, to find the concept descriptions indicating the missing concepts during students' learning. The results of this phase could also be fed back to teachers for refining or reorganizing the learning path.

The rest of this paper is arranged in the following way: Section 3 describes the problem and goal of the presented work. Section 4 introduces the theoretical review of the applied fuzzy clustering techniques. Section 5 introduces the data sets and the preprocessing. Section 6 introduces the experiments design and results analysis. Comparison between the different clustering techniques and the matching with the real world e-learners behaviour and their marks are introduced

.

in section 7. The concluded suggestions and the recommendations are presented in section 8. Finally, the conclusion is outlined in section 9.

### III. PROBLEMS AND GOALS

#### A. Problems

*Web Data Challenges:* Straightforward applications of data mining techniques on web using data face several challenges, which make it difficult to use the statistical clustering techniques. Such challenges as following [29,30]:

§ Data collected during users' navigation are not numeric in nature as traditional data mining.
§ Noise and data incompleteness are important issues for user access data and there are no straightforward ways to handle them.
§ The structure and content of hypermedia systems, as well as additional data, like client-side information, registration data, product-oriented user events, etc., often need to be taken into consideration. Efficiency and scalability of data mining algorithms is another issue of prime importance when mining access data, because of the very large scale of the problems.
§ Statistical measures, like frequency of accessed Web documents, are too simple for extracting patterns of browsing behavior.
§ The users on the Internet are very mobile on the web sites based on their needs and wants.

*The statistical clustering methods are not suitable*[29,30]*:* The statistical clustering provides only the crisp clustering; which does not match with the real world needs, (the real world applications do not consider the world as two halves black and white only).

#### B. Goal of the Work

The goal is the introducing of different fuzzy clustering models, especially the kernelized one as well as the selection of the best model that discovers the students' behavior. Another goal is to overcome the challenges of web usage data.

### IV. THEORETICAL REVIEW OF FUZZY CLUSTERING

One of the main tasks in data mining is the clustering. Clustering is a division of data into groups of similar objects. Each group, called cluster, consists of objects that are similar between themselves and dissimilar to objects of other groups. Representing data by fewer clusters necessarily loses certain fine details, but achieves simplification. Clustering algorithms, in general, are divided into two categories: Hierarchical Methods (agglomerative algorithms, divisive algorithms), and Partitioning Methods (probabilistic clustering, k-medoids methods, k-means methods). Hierarchical clustering builds a cluster hierarchy; every cluster node contains child clusters; sibling clusters partition the points covered by their common parent. Such an approach allows exploring data on different levels of granularity. Hierarchical clustering methods are categorized into agglomerative (bottom-up) and divisive (top-down). An agglomerative clustering starts with one-point (singleton) clusters and recursively merges two or more most appropriate clusters. A divisive clustering starts with one cluster of all data points and recursively splits the most appropriate cluster. The process continues until a stopping criterion (frequently, the requested number k of clusters) is achieved. Data partitioning algorithms divide data into several subsets. Because checking all possible subset possibilities may be computationally very consumptive, certain heuristics are used in the form of iterative optimization. Unlike hierarchical methods, in which clusters are not revisited after being constructed, relocation algorithms gradually improve clusters. The next section describes the theoretical review for the different fuzzy clustering methods used.

#### A. Fuzzy C-Means

Fuzzy clustering is a widely applied method for obtaining fuzzy models from data. It has been applied successfully in various fields. In classical cluster analysis each datum must be assigned to exactly one cluster. Fuzzy cluster analysis relaxes this requirement by allowing gradual memberships, thus offering the opportunity to deal with data that belong to more than one cluster at the same time. Most fuzzy clustering algorithms are objective function based. They determine an optimal classification by minimizing an objective function. In objective function based clustering usually each cluster is represented by a cluster prototype. This prototype consists of a cluster centre and maybe some additional information about the size and the shape of the cluster. The size and shape parameters determine the extension of the cluster in different directions of the underlying domain. The degrees of membership to which a given data point belongs to the different clusters are computed from the distances of the data point to the cluster centers with regard to the size and the shape of the cluster as stated by the additional prototype information. The closer a data point lies to the centre of a cluster, the higher is its degree of membership to this cluster. Hence the problem to divide a dataset into c clusters can be stated as the task to minimize the distances of the data points to the cluster centers, since, of course, we want to maximize the degrees of membership. Most analytical fuzzy clustering algorithms are based on optimization of the basic c-means objective function, or some modification of it. The Fuzzy C-means (FCM) algorithm proposed by Bezdek [31,32] aims to find fuzzy partitioning of a given training set, by minimizing of the basic c-means objective functional as n Eq. (1):

$$f(u, c_1, \ldots, c_C) = \sum_{i=1}^{c} \sum_{j=1}^{n} u_{ij}^m \| x_j - c_i \|^2 \quad (1)$$

Where $u_{ij}$ values are between 0 and 1; $c_i$ is the cluster centre of fuzzy group i, and the parameter m is a weighting exponent on each fuzzy membership. In FCM, the membership matrix U is allowed to have not only 0 and 1 but also the elements with any values between 0 and 1, this matrix satisfying:

$$\sum_{i=1}^{c} u_{ij} = 1, \forall j = 1, \ldots, n \quad (2)$$

Fuzzy partitioning is carried out through an iterative optimization of the objective function shown above, with the update of membership $u_{ij}$ and the cluster centers $c_i$ by:

$$c_i = \frac{\sum_{j=1}^{n} u_{ij}^m x_j}{\sum_{j=1}^{n} u_{ij}^m} \quad (3)$$

$$u_{ij} = \frac{1}{\sum_{k=1}^{c} \left( \frac{\|x_j - c_i\|}{\|x_j - c_k\|} \right)^{-2/(m-1)}} \quad (4)$$

The FCM clustering algorithm steps are illustrated in the following algorithm:
Step 1: Initialize the membership matrix U with random values between 0 and 1 such that the constraints in Equation (2) are satisfied.
Step 2: Calculate fuzzy cluster centers $c_i$, i=1,.., c using Equation (3).
Step 3: Compute the cost function (objective function) according to Equation (1). Stop if either it is below a certain tolerance value or its improvement over previous iteration is below a certain threshold.
Step 4: Compute a new membership matrix U using Equation (4).
Step 5: Go to step 2.
The iterations stops when the difference between the fuzzy partition matrices in two following iterations is lower than $e$.

*B. Kernelized Fuzzy C-Means Method*

The kernel methods [33, 34] are one of the most researched subjects within machine learning community in the recent few years and have widely been applied to pattern recognition and function approximation. The main motives of using the kernel methods consist in: (1) inducing a class of robust non-Euclidean distance measures for the original data space to derive new objective functions and thus clustering the non-Euclidean structures in data; (2) enhancing robustness of the original clustering algorithms to noise and outliers, and (3) still retaining computational simplicity. The algorithm is realized by modifying the objective function in the conventional fuzzy c-means (FCM) algorithm using a kernel-induced distance instead of Euclidean distance in the FCM, and thus the corresponding algorithm is derived and called as the kernelized fuzzy c-means (KFCM) algorithm, which is more robust than FCM. Here, the kernel function K(x, c) is taken as the Gaussian radial basic function (GRBF) as follows:

$$K(x,c) = \exp\left(\frac{-\|x-c\|^2}{s^2}\right) \quad (5)$$

Where $\sigma$: is an adjustable parameter. The objective function is given by

$$f_m = 2 \sum_{i=1}^{c} \sum_{j=1}^{n} u_{ij}^m (1 - K(x_j, c_i)) \quad (6)$$

The fuzzy membership matrix u can be obtained from:

$$u_{ij} = \frac{(1 - K(x_j, c_i))^{-1/(m-1)}}{\sum_{k=1}^{c} (1 - K(x_j, c_i))^{-1/(m-1)}} \quad (7)$$

The cluster center ci can be obtained from:

$$c_i = \frac{\sum_{j=1}^{n} u_{ij}^m K(x_j, c_i) x_j}{\sum_{j=1}^{n} u_{ij}^m K(x_j, c_i)} \quad (8)$$

The proposed KFCM algorithm is almost identical to the FCM, except in step 2, Eq. (8) is used instead of Eq. (4) to update the centers. In step 4, Eq. (7) is used instead of Eq. (3) to update the memberships. The proposed implemented fuzzy clustering including both FCM and KFCM including the post processing technique is shown in Figure 2.

The implemented algorithm consists of two main parts the first is the fuzzy clustering, and the second is the post processing.
The output of the first part will be the U matrix and the cancroids $C_i$, and the outputs of the second part of the algorithm is the fuzzy clusters that consists of the following areas:
1. Area that represent the members in the clusters with high membership values; which called Sure Area (i.e. those members are surly belong to that cluster).
2. The overlapping areas that represent the members; which could not be assigned to any cluster, therefore it will be belong to two or more clusters, this overlapping area called the May Be Areas. These areas may help in taking a decisions as; the sure areas says that those elements are surly belong to those clusters, as well as the May Be Areas also says that; those elements are not be essential in taking a decisions.
Another benefit of the overlapping areas is how to focus on the overlapping areas between specific two clusters; that can help in the study of how to attract the students from one class to another.

V. DATA SETS AND DESIGN OF THE EXPERIMENT

*A. Log Files Description*

The data recorded in server logs reflects the access of a Web site by multiple users. Web server-side data and client-side data constitute the main sources of data for Web usage mining. Web server access logs constitute the most widely used data because it explicitly records the browsing behavior of site visitors. For this reason, the term Web log mining is sometimes used. Web log mining should not be confused with Web log analysis. An illustrative example for the log file is shown in Table 1.

*B. Data Set Description*

The data sets used in this study were obtained from web access logs for studying a two courses; the first is for teaching "data structures"; the course is offered in the second term of the second year, at computing science programme at Saint Mary's University. The second course is "Introduction to Computing Science and Programming", for the first year. Data were collected over 16 weeks (four months). The number of students in these courses is described in details in Table 2. From the work presented in [29-30], the student's behavior through teaching courses it proposed that, visits from

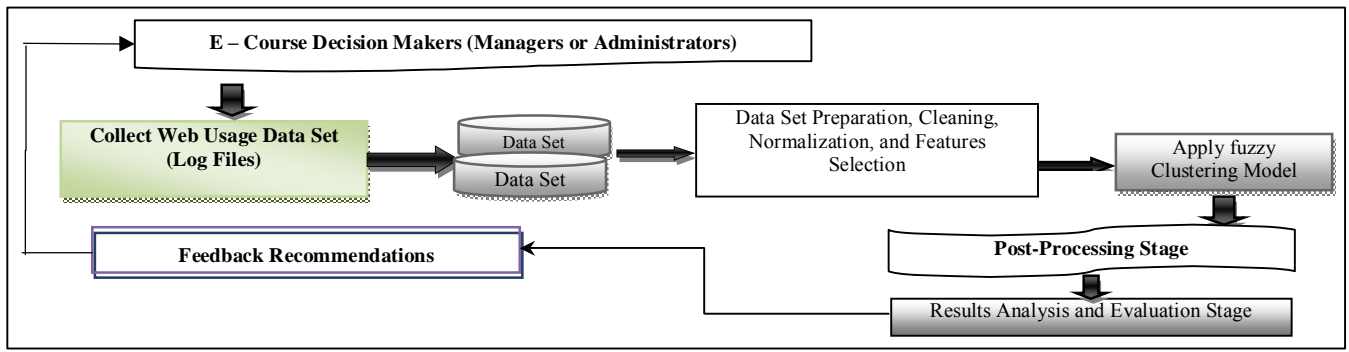

**Figure1. The proposed applied data mining system**

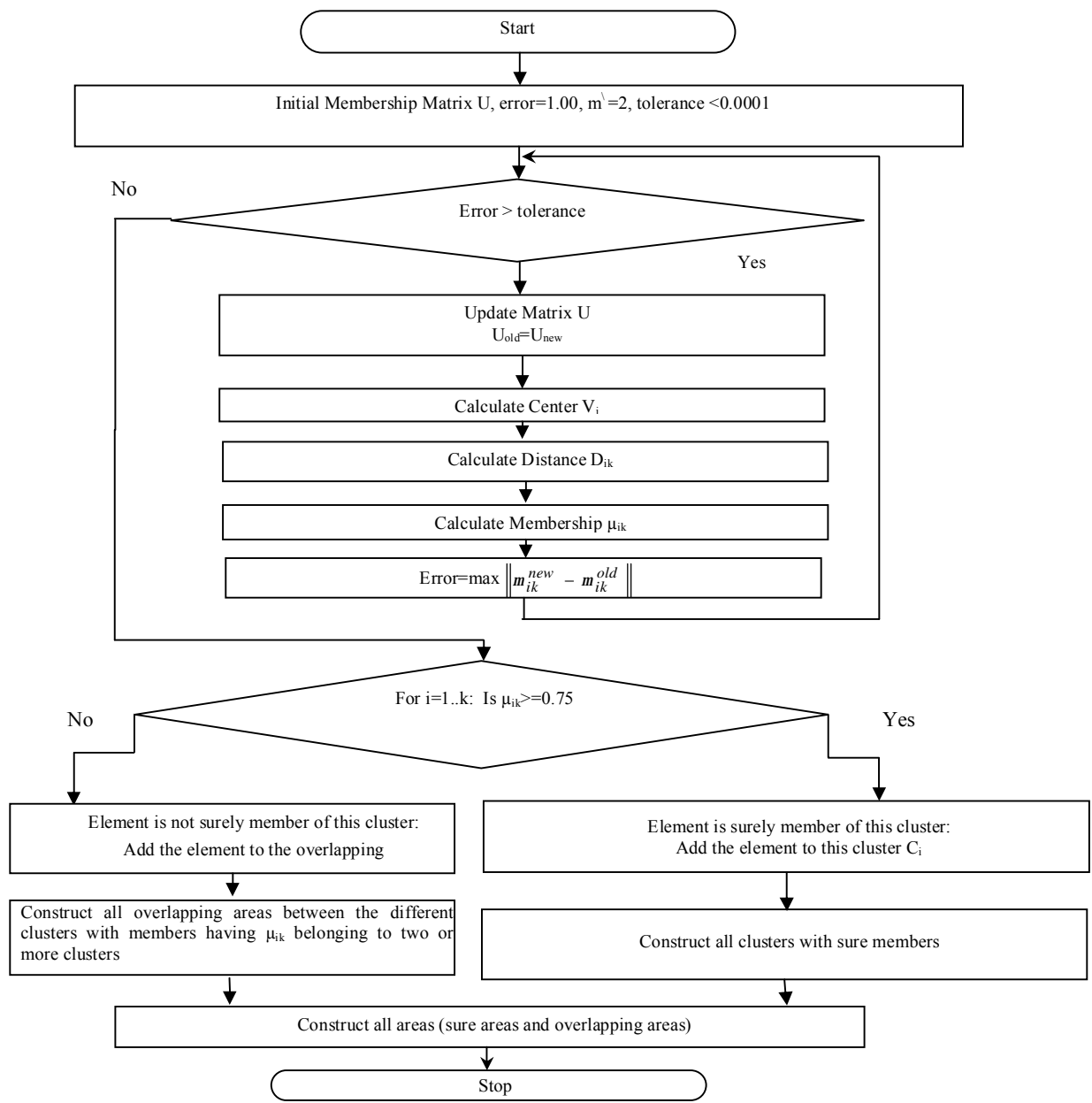

Figure2. The proposed Clustering Models (FCM and KFCM) and the Post-processing Technique

students attending this course could fall into one of the following five categories:

1. Regular students: These learners download the current set of notes. Since they download a limited/current set of notes, they probably study class-notes on a regular basis.
2. Bad students: These learners download a large set of notes. This indicates that they have stayed away from the class-notes for a long period of time. They are planning for pretest cramming.
3. Worker students: These visitors are mostly working on class or lab assignments or accessing the discussion board.
4. Casual students: those students who did not interact with the course material and if they visit the web course, they do not download any documents.
5. Absent students: those students who are absent during the teaching course.

Where after many experiments we found that the casual students and the absent students do not affect the study of learner's profiles because the paper focuses on the learners profiles based on number of hits, downloaded documents, time of accessing the web course, and day of accessing the course materials.

### C. Data Preparation and Cleaning

Data quality is one of the fundamental issues in data mining. Poor quality of data always leads to poor quality of results. Sometimes poor quality data results in interesting, or unexpected results. Therefore data preparation is a crucial step before applying data mining algorithms. In this work data preparation; consists of two phases, data cleaning, and data abstraction and normalization.

1. Data cleaning process:Data cleaning process consists of two steps Hits cleaning and Visits cleaning as following:

• *Hits Cleaning:* To remove the hits from search engines and other robots. In the second data set; the cleaning step reduced the log files data set by 3.5%, the number of hits was reduced from 40152 before cleaning to 36005 after cleaning.

• *Visits cleaning:* To clean the data from those visits, which didn't download any class-notes, were eliminated, since these visits correspond to casual visitors. The total visits were 4248; after visits cleaning the visits were reduced to 1287 as shown in Table 3.

• *Remove the Casual and absent classes from the data sets:* where those two cleaning steps were not interested in studying the learners who did not download any Byte, as well as the casual learners.

• *Data privacy and learners security:* It is required for the identification of web visits; it is done using Linux commands. Certain areas of the web site were protected, and the users could only access them using their IDs and passwords. The activities in the restricted parts of the web site consisted of submitting a user profile, changing a password, submission of assignments, viewing the submissions, accessing the discussion board, and viewing current class marks. The rest of the web site was public. The public portion consisted of viewing course information, a lab manual, class-notes, class assignments, and lab assignments. If the users only accessed the public web site, their IDs would be unknown. Therefore, the web users were identified based on their IP address. This also made sure that the user privacy was protected. A visit from an IP address started when the first request was made from the IP address. The visit continued as long as the consecutive requests from the IP address had sufficiently small delay. The web logs were preprocessed to create an appropriate representation of each user corresponding to a visit.

2. Data Abstraction and Normalization:The abstract representation of a web user is a critical step; that requires a good knowledge of the application domain. Previous personal experience with the students in the course suggested that some of the students print preliminary notes before a class and an updated copy after the class. Some students view the notes on-line on a regular basis. Some students print all the notes around important dates such as midterm and final examinations. In addition, there are many accesses on Tuesdays and Thursdays, when the in-laboratory assignments are due. On and Off-campus points of access can also provide some indication of a user's objectives for the visit. Based on some of these observations, it was decided to use the following attributes for representing each visitor:

a. On campus/Off campus access (binary values 0 or 1).
b. Day time/Night time access: 8 a.m. to 8 p.m. were considered to be the Daytime (day/night).
c. Access during lab/class days or non-lab/class days: All the labs and classes were held on Tuesdays and Thursdays. The visitors on these days are more likely to be Worker Students.
d. Number of hits (decimal values).
e. Number of class-notes downloads (decimal values).

The first three attributes had binary values of 0 or 1. The last two values were normalized. The distribution of the number of hits and the number of class-notes was analyzed for determining appropriate weight factors. The numbers of hits were set to be in the range [0, 10]. Since the class-notes were the focus of the clustering, the last variable was assigned higher importance, where the values ranged from 0 to 15. Even though the weight for class-notes seems high, the study of actual distributions showed that 99% of visits had values less than 15 for the data set.

### VI. EXPERIMENTS DESIGN AND RESULTS ANALYSIS

It was possible to classify the learners using the two fuzzy clustering techniques into five clusters as regular students, worker students, bad students, casual students, and absent students using both of fuzzy c-means, kernelized c-means, and KFCM Method. But the problem here is that the absent students were not found in the data sets as the absent student is characterized by the casual interaction with the web course, they did not download any materials documentation related to the course when they visited the web site.

Table 1: Common Log File Format

EXAMPLE: 24.138.46.172--[09/AUG/2001:20:52:07-0300] GET/~CSC226/PROJECT1.HTMHTTP/1.1 200 4662

| FIELD IN THE LOG FILE RECORD | VALUE |
|---|---|
| Client IP address or hostname (if DNS lookups are performed) | 24.138.46.172 |
| Client's username (if a login was required), or "--" if anonymous | -- |
| Access Date | 09/AUG/2001 |
| Access Time | 20:52:07-0300 |
| HTTP request method (GET, POST, HEAD.) | GET |
| Path of the resource on the Web server (identifying the URL) | ~CSC226/PROJECT1HTM |
| The protocol used for the transmission (HTTP/1.0, HTTP/1.1) | HTTP/1.1 |
| Service status code returned by the server (200 for OK, and 404 not found) | 200 |
| Number of bytes transmitted | 4662 |

Table 2: Historical Description of the Courses

| Course | | Description |
|---|---|---|
| Introduction to Computing Science and Programming for First year in First term | | The initial number of students in the course was 180. The number changed over the course of the semester to 130 to 140 students. Students in the course come from a wide variety of backgrounds, such as Computing Science major hopefuls, students taking the course as a required science course, and students taking the course as a science or general elective. |
| Data structures Second year in second term | | The number of students in this course was around 25 students and the number changed to 23 students. This course was more difficult but the students were more stable in this course. |

Table 3: Data Sets Before and After Preprocessing

| Data Set | Hits | Hits After Cleaning | Visits | Visits After Cleaning |
|---|---|---|---|---|
| First Course Data Set | 361609 | 343000 | 23754 | 7673 |
| Second Course Data Set | 40152 | 36005 | 4248 | 1287 |

Table 4: FCM Results for 1st Data Set

| Class Name | Behavior of each Class | | | | | Size |
|---|---|---|---|---|---|---|
| | Camp. | Time | Lab | Hits | Req. | |
| **Regular** | 0.002 | 0.65 | 0.34 | 0.49 | 0.70 | 1904 |
| **Workers** | 0.98 | 0.92 | 0.66 | 0.98 | 1.2 | 2550 |
| **Bad** | 0.67 | 0.732 | 0.45 | 3.23 | 6 | 396 |
| **R&W** | 0.22 | 0.68 | 0.42 | 0.53 | 0.8 | 2600 |
| **R&B** | 0.3 | 0.68 | 0.38 | 2 | 2.8 | 98 |
| **W&B** | 0.77 | 0.81 | 0.53 | 1.03 | 1.01 | 125 |
| **R&W&B** | 0.45 | 0.72 | 0.39 | 0.37 | 0.99 | 98 |

Table 5: KFCM Results for 1st Data Set

| Class Name | Behavior of each Class | | | | | Size |
|---|---|---|---|---|---|---|
| | **Camp.** | **Time** | **Lab** | **Hits** | **Req.** | |
| **Regular** | 0.006 | 0.58 | 0.44 | 0.38 | 0.77 | 1870 |
| **Workers** | 1 | 0.78 | 0.59 | 1 | 1.4 | 2430 |
| **Bad** | 0.7 | 0.65 | 0.35 | 4 | 6.5 | 416 |
| **R&W** | 0.3 | 0.53 | 0.49 | 0.8 | 0.9 | 2654 |
| **R&B** | 0.39 | 0.49 | 0.22 | 2 | 3 | 78 |
| **W&B** | 0.82 | 0.72 | 0.28 | 3 | 0.9 | 225 |
| **R&W&B** | 0.47 | 0.59 | 0.37 | 0.3 | 1.22 | 78 |

Table 6: FCM Results for 2nd Data Set

| Class Name | Behavior of each Class | | | | | Size |
|---|---|---|---|---|---|---|
| | Camp. | Time | Lab | Hits | Req. | |
| Regular | 0.48 | 0.65 | 0.31 | 2.08 | 3.99 | 161 |
| Workers | 0.54 | 0.70 | 0.42 | 2.40 | 2.75 | 1000 |
| Bad | 0.57 | 0.55 | 0.45 | 2.24 | 4.84 | 25 |
| R&W | 0.54 | 0.75 | 0.51 | 0.90 | 2.9 | 54 |
| R&B | 0.58 | 0.74 | 0.51 | 0.94 | 4 | 47 |
| W&B | - | - | - | - | - | 0 |
| R&W&B | - | - | - | - | - | 0 |

Table 7: KFCM Results for 2nd Data Set

| Class Name | Behavior of each Class | | | | | Size |
|---|---|---|---|---|---|---|
| | Camp. | Time | Lab | Hits | Req. | |
| Regular | 0.42 | 0.55 | 0.39 | 2.3 | 3 | 168 |
| Workers | 0.64 | 0.74 | 0.46 | 2.7 | 2.2 | 977 |
| Bad | 0.68 | 0.6 | 0.33 | 3.1 | 4.3 | 49 |
| R&W | 0.50 | 0.75 | 0.51 | 0.90 | 0.58 | 50 |
| R&B | 0.54 | 0.74 | 0.51 | 0.94 | 0.74 | 43 |
| W&B | - | - | - | - | - | 0 |
| R&W&B | - | - | - | - | - | 0 |

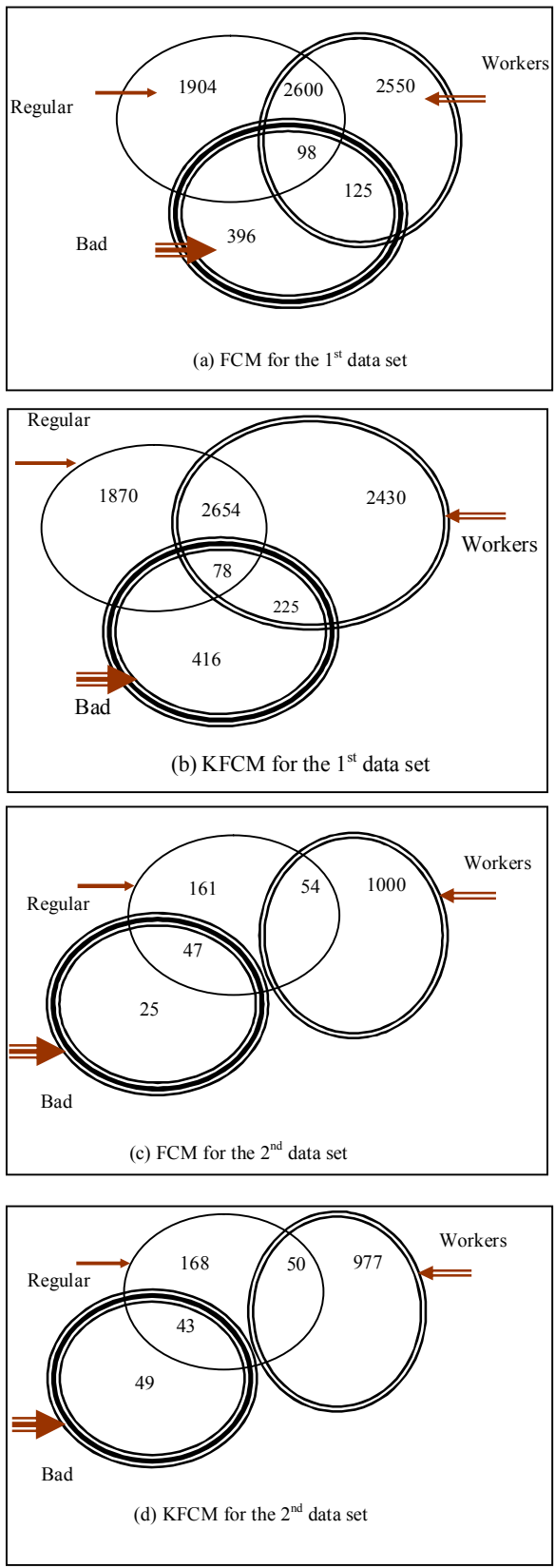

Figure3. Fuzzy Clusters Representation: a and b for the 1st Data Set, and c and d for the 2nd Data Set

Table 8: Comparison Between Results of FCM and KFCM

| Data Set | Model | Ratio Between Size of Clusters and Real Results | | |
|---|---|---|---|---|
| | | Regular /Real | Worker /Real | Bad /Real |
| 1st | FCM | 81% | 88% | 90% |
| 1st | KFCM | 87.5 | 91% | 93% |
| 2nd | FCM | 88% | 90% | 96% |
| 2nd | KFCM | 88.07 | 90.9% | 98% |

Table 9: Questionnaires Results for the Second Course

| Students % | Students' Opinion |
|---|---|
| 21.739% | Accept online only as interactive method |
| 17.39% | Refused on-line as a method not usual |
| 30.43% | Hybrid of on-line and printed document |
| 21.739% | Refused on-line (not used to work with it) |
| 6.69% | Refused on line (due to practical reasons) |

Therefore we have decided to re-cluster the data sets into three clusters only as regular, workers, and bad students; and neglect both the absent and casual students classes. The results were good enough to reflect the learner's behaviour on the e-course. Table 6 and Table 7 show details of clusters for the 2nd data set. Each cluster is characterized by the following:

1. The number of Bad Students was significantly less than the numbers of Worker Students and Regular Students visitors, and Bad Students class was identified by the high number of hits and document-downloads.
2. The size of the Worker Students class was the biggest one, and identified by the lowest number of hits and document-downloads.
3. The size of the Regular Students class was moderate smaller than Worker Students and larger than Bad Students, identified by the moderate number of hits and document-downloads, and regularity of downloading behaviour.

The interpretation of the results obtained from this phase is as same as the interpretation for the results of the first data set shown in Table 4, and Table 5. The fuzzy representation of the clustering results for the different clusters and their overlapping are presented in Figure3.

VII. COMPARISON ANALYSIS BETWEEN FCM & KFCM

Both FCM and KFCM were able to cluster the data sets as shown in Tables 4,5,6,7 and Figure3 (a), (b), (c), (d), with moderate accuracy. Moreover, the results obtained from KFCM were better when compared with the real marks of the students and the ratios of the students with different grades, the calculations were done as a ratio, for example the majority of the grades were grad B+ that can fit with workers students, the next large grade was the A that match with the class regular, finally the minority that has grade C and fall in the course (grade D) was matched with the Bad student class. Table 8 illustrates the matching between the obtained results

-

from FCM and KFCM for the two data sets, and the real marks and grades of the students. The comparison concludes that both of the two methods were good enough; moreover the KFCM was better and its performance from the points of matching with the real marks and the speed was high.

## VIII. SUGGESTIONS AND RECOMMENDATIONS

### A. Student feedback

Feedback from students on the second course indicates that; there are some concerns over accessing, reading internet pages, and downloading different materials as shown in Table 9. From the table we conclude some points as following:

1. Due to practical reasons such as eye strain, portability, navigation and the process of developing understanding by adding notes.
2. Students have opinions that the materials were difficult and not have more explanations, others said that the course itself is more difficult to follow on-line; they thought that it is a difficulty added to the course itself.
3. Students suggested that the combination of online and printed versions of materials would be better.
4. Students were satisfied using on-line more than the off-line; as it gives the feeling of the classroom environment.
5. Students raise the need to make it easier to obtain print versions for easier handling process because of their usual (not because it is difficult but because they did not used to).

### B. The Suggestions and Recommendations:

1. Formative evaluation: It is the evaluation of an educational program while it is still in development, and with the purpose of continually improving the program. Examining how students use the system is one way to evaluate the instructional design in a formative manner and it may help the educator to improve the instructional materials.
2. Oriented towards students: The objective is to recommend to learners activities, resources and learning tasks that would favor and improve their learning, suggest good learning experiences for the students, suggest path pruning and shortening or simply links to follow, based on the tasks already done by the learner and their successes, and on tasks made by other similar learners, etc.
3. Oriented towards educators: The objective is to get more objective feedback for instruction, evaluate the structure of the course content and its effectiveness on the learning process, classify learners into groups based on their needs in guidance and monitoring, find learning learner's regular as well as irregular patterns, find the most frequently made mistakes, find activities that are more effective, discover information to improve the adaptation and customization of the courses, restructure sites to better personalize courseware, organize the contents efficiently to the progress of the learner and adaptively constructing instructional plans, etc.
4. Oriented towards academics responsible and administrators: The objective is to have parameters about how to improve site efficiency and adapt it to the behavior of their users (optimal server size, network traffic distribution, etc.), have measures about how to better organize institutional resources (human and material) and their educational offer, enhance educational programs offer and determine effectiveness of the new computer mediated distance learning approach. There are many general data mining tools that provide mining algorithms, filtering and visualization techniques. Some examples of commercial and academic tool are DBMiner, Clementine, Intelligent Miner, Weka, etc. As a total conclusion, the suggestions and the recommendations from this work are focused on: the educators' behavior obtained from both fuzzy clustering models.

## IX. CONCLUSIONS

The work presented in this paper focuses on how to find good models for the evaluation for E-learning systems. The paper introduces the use of two different fuzzy clustering techniques, the FCM and the KFCM, where the clustering is one of the most important models in data mining. Both of FCM and KFCM clustering were able to find the clusters for the learners and the results were matched with the real marks of the students with high percentage. Moreover the KFCM results have high matching percentage with the real marks than the FCM. The suggestion and the recommendations were constructed based on the clustering results and the questioners obtained from the students that represent the learners' profiles and reflect their behavior during the teaching of the e-course. Finally the paper proved that the ability of fuzzy clustering generally and KFCM was better in predicting the e-learners behaviour.

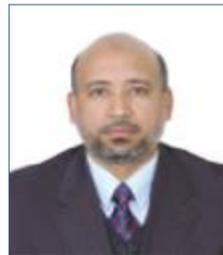

**Mofreh A. Hogo** is a lecturer at Benha University, Egypt. He is a lecturer of Computer Science and Engineering. Dr. Hogo holds a PhD in Informatics Technology from Czech Technical University in Prague, Computer Science and Engineering Dept. 2004. He is the author of over 40 papers that have been published in refereed international Journals (Information Sciences, Elsiver, UBICC, IJICIS, IJCSIS, IJPRAI, ESWA, IJEL, Web Intelligence and Agent Systems, Intelligent Systems, international journal of NNW, IJAIT Journal of Artificial Intelligence Tools, IJCI) and Book chapters (Neural Networks Applications in Information Technology and Web Engineering Book, Encyclopedia of Data Warehousing and Mining, and Lecture Notes in Artificial Intelligence Series), and international conferences (Systemics, Cybernetics and Informatics Information Systems Management, IEEE/WIC, IEEE/WIC/ACM, ICEIS). His areas of interest include Digital Image Processing, Multimedia Networks, Intrusion detection, Data Mining, Data Clustering and classification, pattern Recognition, character recognition, fuzzy clustering, artificial Neural Networks, Expert systems, Software Engineering.